# Source-like solution for radial imbibition into a homogeneous semi-infinite porous medium

*Junfeng Xiao[1], Howard A. Stone[2] and Daniel Attinger[3]\**

[1]Columbia University, Department of Mechanical Engineering, New York NY 10027

[2]Princeton University, Department of Mechanical and Aerospace Engineering, Princeton NJ 08544

[3]Iowa State University, Department of Mechanical Engineering, Ames IA 50011

January 23, 2012

ABSTRACT

We describe the imbibition process from a point source into a homogeneous semi-infinite porous material. When body forces are negligible, the advance of the wetting front is driven by capillary pressure and resisted by viscous forces. With the assumption that the wetting front assumes a hemispherical shape, our analytical results show that the absorbed volume flow rate is approximately constant with respect to time, and that the radius of the wetting evolves in time as $r \approx t^{1/3}$. This cube-root law for the long-time dynamics is confirmed by experiments using a packed cell of glass microspheres with average diameter of 42 μm. This result complements the classical one-dimensional imbibition result where the imbibition length $\ell \approx t^{1/2}$, and studies in axisymmetric porous cones with small opening angles where $\ell \approx t^{1/4}$ at long times.

---

\* Corresponding Author. E-mail: attinger@iastate.edu

INTRODUCTION

When a liquid contacts a dry porous medium or a capillary tube with a wetting angle smaller than π/2, it is transported by capillary forces into the medium or the tube. This motion is resisted by viscous forces. In a one-dimensional geometry such as a porous rectangular strip or a capillary tube, the wetting front advances according to a power law $\ell^2 = D\,t$, with $\ell$ the distance, $D$ a coefficient that depends on the porous medium and the liquid, and $t$ the time. This power law is best known as the 'Lucas-Washburn law'[1], and is also described as 'diffusive imbibition'[2]. As reviewed in[3], the phenomenon of diffusive imbibition is of importance for fluid transport in plants and soils, and in industrial processes such as printing, oil recovery, cooking, wine filtering, fabrication of composite materials, behavior of garments. The forensic discipline of bloodstain pattern analysis would also benefit from a better understanding of stain formation on clothes or carpets for example[4]. Imbibition processes can also play a crucial role in low-cost microfluidic devices made of paper[5].

Recently, imbibition has been studied for geometries with more than one dimension. For example, Clarke et al.[6], Anderson[7] and Hilpert and Ben-David[8] modeled the imbibition of a finite-size droplet into a porous medium, accounting for the deformation of the drop while assuming that the pores fill in the normal direction to the surface of the porous medium. Oko et al.[9] used high-speed imaging to measure the imbibition and evaporation of picoliter water droplets on paper media commonly used for inkjet printing. Reyssat *et al*[2] studied the imbibition in a cone geometry with small opening angle α, and showed that at later times the imbibition varied as $\ell \approx t^{1/4}$. Mason et al. studied analytically imbibition towards the center of porous cores with cylindrical, spherical and toroidal geometries[10]. Also, Mendez *et al*[11] described the two-dimensional imbibition process in a thin porous membrane with a fan shape, specifically a rectangular sector attached to a circular sector, and expressed the deviation from the Lucas-Washburn law in the circular sector. In this paper, we describe the imbibition process from a point source into a homogeneous semi-infinite porous material.

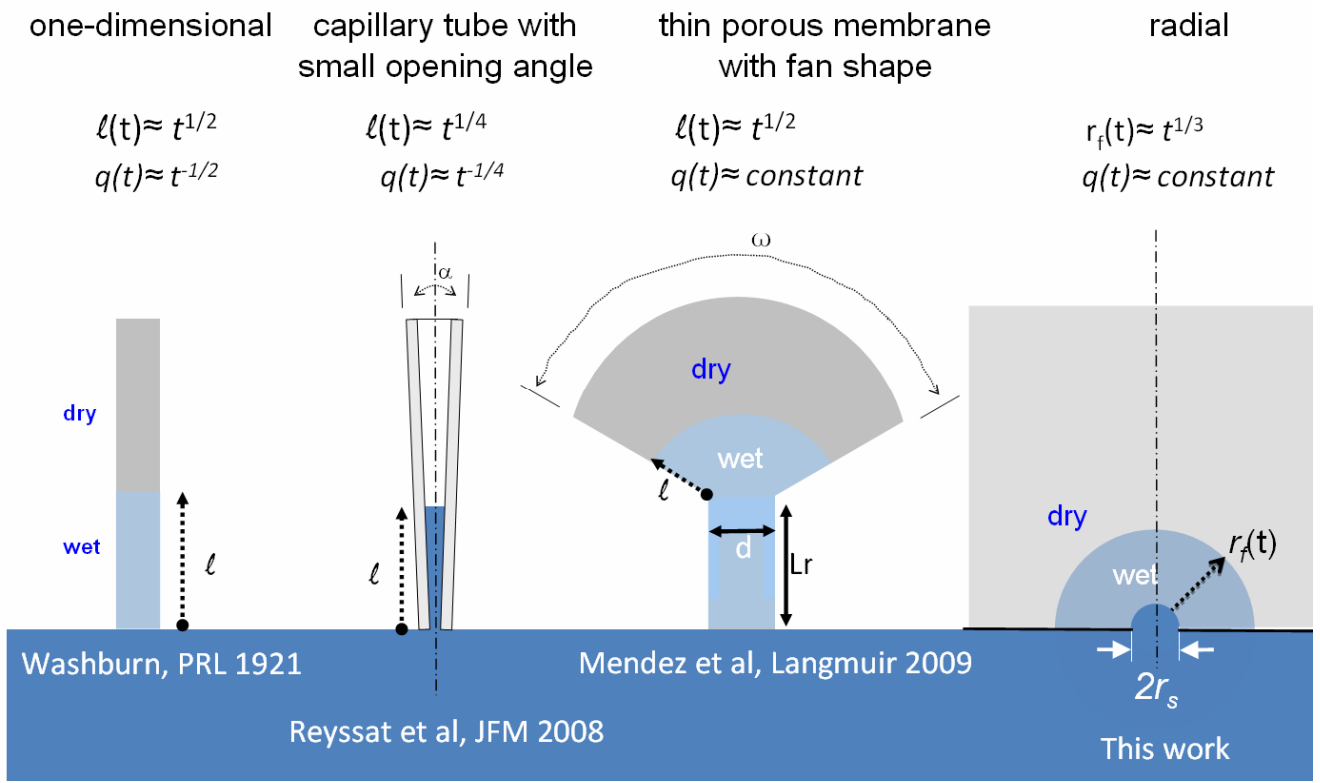

**Figure 1.** From left to right, four analytical solutions for imbibition in porous media, as a function of the geometry: one dimensional imbibition process that obeys the Lucas-Washburn law[1a]; imbibition in a capillary tube with small opening angle[2], in a thin, fan-shape membrane[11], and in the radial geometry considered in this work. For each case the evolution of the wetted length $\ell$ and flow rate $q$ (at long times) are expressed as a power law of the time $t$.

ANALYTICAL MODEL

In this section, we derive a mathematical model for the imbibition from a point source into a semi-infinite geometry illustrated in Figure 2. The flow in a porous media is described by Darcy's law:

$$\mathbf{v} = -\frac{k}{\mu}\nabla p, \qquad (1)$$

where $k$, $\mu$ and **v** respectively represent the permeability (m$^2$) of the medium, dynamic viscosity (Pa·s) of the liquid and the average velocity. Here the porous material is assumed to be homogeneous and isotropic, so that the permeability $k$ is a scalar. If liquid loss due to evaporation from the porous media can be neglected, mass conservation for an incompressible flow yields

$$\nabla \cdot v = 0, \text{ so that } \nabla^2 p = 0. \tag{2}$$

The capillary pressure at the advancing front of the wetted region is determined by the Laplace pressure:

$$p_c = \frac{2\gamma \cos\theta_c}{r_p}, \tag{3}$$

where $\gamma$, $r_p$ and $\theta_c$ are, respectively, the surface tension between the liquid and the atmosphere, a representative radius of the pores, and the contact angle of liquid on the porous material.

Several standard assumptions are made in the derivation below:

1. The source of liquid is small enough to be considered as a point source, which can provide an infinite liquid supply for the imbibition process.
2. The shape of the advancing liquid front is a hemisphere, so that liquid flows radially from the source to the advancing front and the radial velocity is uniform along the advancing front. The advancing wetting front of liquid is then defined in spherical coordinates as $r = r_f(t)$, with the pressure gradient only radial.
3. The effect of gravity on the liquid flow is neglected, which means hydrostatic pressure is much smaller than capillary pressure.

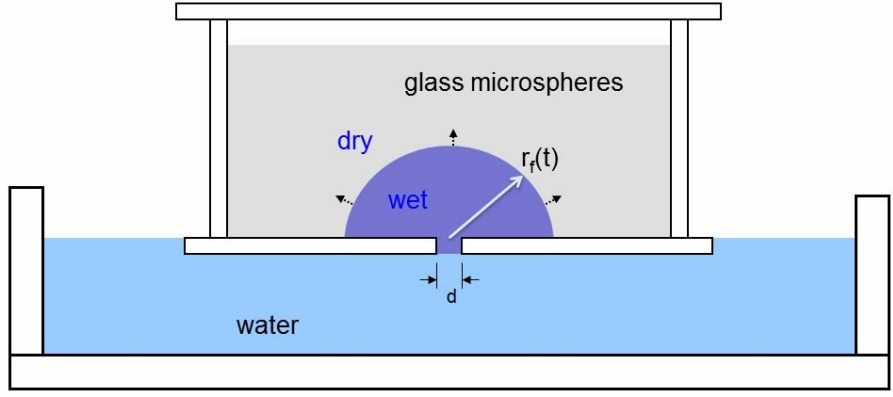

**Figure 2.** An axisymmetric geometry for the hemispherical imbibition process in a semi-infinite porous medium. Liquid is withdrawn from a large reservoir through a hole with diameter $d$.

The volumetric flow rate of liquid $q(t)$ at the advancing front is:

$$q(t) = 2\pi r_f^2 u_r, \qquad (4)$$

where $u_r = \dfrac{dr_f}{dt}$ denotes the radial velocity of liquid at the advancing front. The radial velocity can be expressed as a function of the pressure gradient using Darcy's law (1):

$$-\frac{dp}{dr} = \frac{\mu q(t)}{2\pi r^2 k}. \qquad (5)$$

The pressure at *the source* is $p_{atm}$, while the pressure at the wetting front $r = r_f$ is $p_{atm} - p_c$. Equation (5) can be integrated from the source at $r \approx r_s$ to the front $r = r_f(t)$ in order to obtain for the pressure

$$p_c = \frac{\mu q(t)}{2\pi k}\left(\frac{1}{r_s} - \frac{1}{r_f}\right). \qquad (6)$$

Combining (6) with (4) yields

$$r_f^2 \frac{dr_f}{dt}\left(\frac{1}{r_s} - \frac{1}{r_f}\right) = \frac{kp_c}{\mu}. \qquad (7)$$

This first-order nonlinear ordinary differential equation for the position of the wetting front $r_f$ can be integrated directly to obtain, with $r_f = r_s$ as the initial condition,

$$\frac{1}{3r_s}\left(r_f^3 - r_s^3\right) - \frac{1}{2}\left(r_f^2 - r_s^2\right) = \frac{kp_c t}{\mu}. \tag{8}$$

At long times, equation (8) simplifies to

$$r_f = \left(\frac{3kp_c r_s}{\mu}\right)^{\frac{1}{3}} t^{\frac{1}{3}}. \tag{9}$$

According to equation (4), in this long time limit, the flow rate $q(t)$ is independent of time, expressed as

$$q = \frac{2\pi k p_c r_s}{\mu}. \tag{10}$$

Equation (9) can be recast in a non-dimensional manner:

$$R(T) = T^{\frac{1}{3}}, \tag{11}$$

by choosing dimensionless parameters as $R = \dfrac{r_f}{r_s}$ and $T = \dfrac{3kp_c}{\mu r_s^2} t$.

EXPERIMENTAL SECTION

Experiments were conducted to measure the evolution of a wetting front with respect to time. Soda lime glass microspheres (P2043SL, Cospheric LLC) were loaded in a polycarbonate box (60 mm x 60 mm x 60 mm) to form a porous medium in the experiments. The microspheres were well mixed in the box using an orbital shaker (miniRoto S56, Fisher Scientific) before each experiment to avoid inhomogeneities in the porous medium. To measure the particle size, the microspheres were immobilized in water between a microscope slide and cover slide, and imaged under an inverted microscope. A MATLAB code was developed to determine the size of 3000 microspheres: the average diameter $d_m$ was measured to be 42 μm, with standard deviation $\sigma = 7$ μm (see Figure 3). The porosity was measured to be $0.36 \pm 0.02$, by comparing three times the weight of an 8 mL beaker filled with dry sand to the same system wetted with water. This porosity value is slightly lower than values measured for packed beds of same size glass particles, as e.g. in[12] where particles with diameter of 40 μm were measured to have average porosity of 0.45. The porosity value found in our study is very close however to the porosity of 0.38 ± 0.02 measured in packed beads[13] with grains size between 40 and 300 μm.

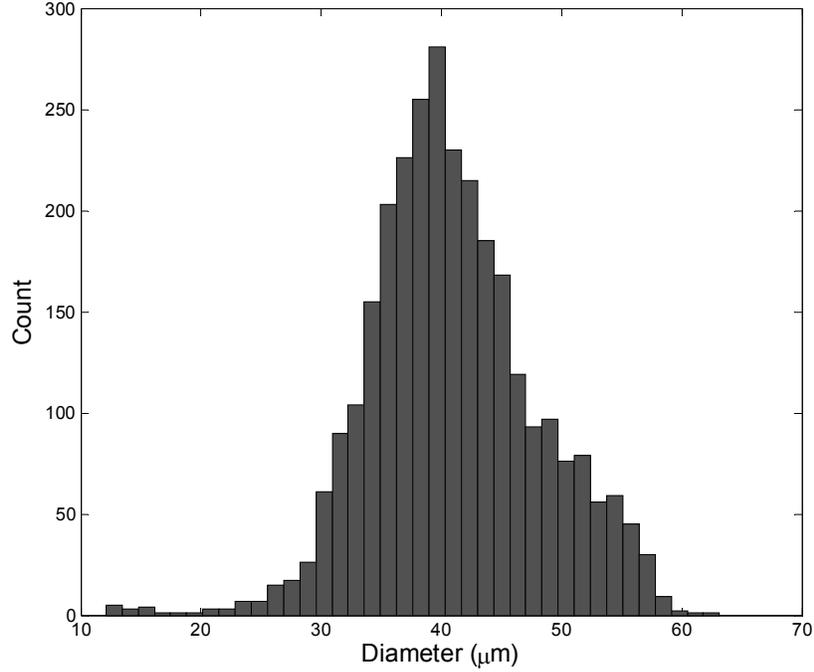

**Figure 3.** The measured size distribution of glass microspheres used in the experiments.

Two important parameters in equation (9) need to be determined, namely the permeability $k$ and capillary pressure $p_c$. For a random packing of microspheres of uniform size, the permeability $k$ of the porous media, as well as the porosity $\varepsilon$, can be determined numerically from the mean diameter of the microspheres $d_m$ using simulations of random sequential deposition events, as shown in[14]. The permeability $k$ of the porous media can also be related to the diameter of the microspheres $d_m$, and the porosity $\varepsilon$ estimated using the Kozeny-Carman model[15]:

$$k = \frac{d_m^2}{180} \frac{\varepsilon^3}{(1-\varepsilon)^2}. \tag{12}$$

We measured the permeability experimentally. A water flow was driven vertically through a Teflon tube (inner diameter 1.58 mm, length 100 mm) filled with glass microspheres, with the flow rate controlled by a syringe pump. A thin metal mesh membrane (15 μm, Betamesh, BOPP AG, Switzerland) with negligible pressure drop was attached to the exit of the tube to retain the microspheres. The pressure difference in the porous media was measured by a pressure gauge (DPG 110, Omega). Then the permeability of the porous media can be calculated from the measured pressure difference and flow rate according to Darcy's law. The permeability was measured to be $k = (1.21 \pm 0.07) \times 10^{-12}$ m$^2$. The experimental value of permeability is in good agreement with the value calculated using the Kozeny-Carman model[15] $k = 1.12 \times 10^{-12}$ m$^2$, or determined numerically in[14] as $k = 9.5 \times 10^{-13}$ m$^2$.

The capillary pressure was measured by a 1-D capillary rise experiment, where vertical imbibition is resisted by gravity, until an equilibrium height $h$ is reached where the hydrostatic pressure is balanced by the capillary pressure, with

$$\rho g h = 2\gamma \cos\theta_c / r_p. \tag{13}$$

The final height $h$ obtained experimentally corresponds to a capillary pressure of 3480 Pa +/-70 Pa. That value was used as the capillary pressure in equations (8) and (9), and in the results shown in Figure 4. If we assume, as in[13], that the largest pore size, i.e. the maximum radius of curvature, controls the capillary pressure, we can obtain independent values for $r_p$ and $\theta_c$ as follows. The capillary pressure obtained by the capillary rise experiment corresponds by equation (13) to a maximum radius of curvature $\cos\theta_c/r_p = 2.39 \times 10^4$ m$^{-1}$. Neglecting the non-uniformity of the beads, we assume that all beads with diameter $d_m$ are packed in a cubic face-centered arrangement, with unit cell distance $\sqrt{8}d_m/2$. From simple analytical geometry considerations, a relation can be established between the wetting angle and the largest radius of curvature. Iterations between that relation and equation (13) converge towards values of a maximum radius of curvature of 33μm, which corresponds to a maximum pore size of 26μm and a wetting angle of 38°, respectively. These estimates are compatible with the actual bead radius (21 μm) and with published literature[16], where the wetting angle of water on soda lime glass was measured as 31°. Note that this method of determining independent values of $r_p$ and $\theta_c$ is only provided as a side note, since the interpretation of the radial imbibition experiments studied in this paper only relies on the value of the capillary pressure measured in the 1-D capillary rise experiment.

The experimental setup to measure the dynamics of the imbibition process is illustrated in Figure 2. A hole with diameter $d = 0.64$ mm and length of 5.9 mm was drilled in the bottom plate (thickness 5.88 mm) as the inlet for the liquid. In the experiments, the height of the wet volume was always kept below 30 mm, which is a height corresponding to less than 10% of the capillary pressure generated by 42 μm spheres. The box loaded with microspheres was then put in a water-filled glass Petri dish, and the liquid level in the Petri dish was adjusted to be the same as the bottom surface of the polycarbonate box, which assures that pressure at the inlet is atmospheric and that the imbibition process has access to a nearly infinite liquid supply. The liquid inlet hole was filled with microspheres, so that the wetting liquid reached the top side of the box bottom in less than 5 seconds, as measured by filling the box with a very thin layer of sand. This estimate is in good agreement with predictions from the Lucas-Washburn law, which estimates the invasion time to be less than 1 second. Experiments with fluids that fully wets glass (silicone oils) were inconclusive because spreading proceeded faster along the plastic container walls than in the sand and produce a flat pancake-shaped wetted region in the sand.

The box was capped during the experiments to avoid evaporation. In order to determine the time dependence of the imbibition front the wet region was measured at different time intervals. In particular, after a given time interval, all of the dry microspheres were poured out of the box by gravity[17], by suddenly turning the box upside down. Microspheres in the wet volume stuck to each other due to capillary adhesion caused by water between the spheres. The wet region also adhered to the bottom plate of polycarbonate box, The shape of the wet volume (photographed in Figure 4), represented by the height ($z_e(t)$) and radius in the orthogonal directions ($r_L(t)$ and $r_R(t)$), was then measured using a caliper. The experiment was repeated with new dry microspheres for another time interval and in this way the time evolution of the advancing front is determined. This procedure was repeated five times in order to have five values of $z_e(t)$, $r_L(t)$ and $r_R(t)$ for each of the six times reported in Figure 4. The experimental error for the imbibition measurement is shown also. Errors were mainly due to the uncertainty of the dynamic viscosity (+/- 7%), and repeatability and uncertainty in the length measurement of the wet region (+/- 0.7mm). The latter absolute error is the reason why vertical error bars at earlier times are larger.

RESULTS AND DISCUSSION

This section compares the theoretical predictions, equations (8) and (9), for imbibition from a point source in a semi-infinite domain to the results of our experiments. To plot the theoretical curve in Figure 4, the capillary pressure is estimated according to the 1-D capillary rise measurement described above.

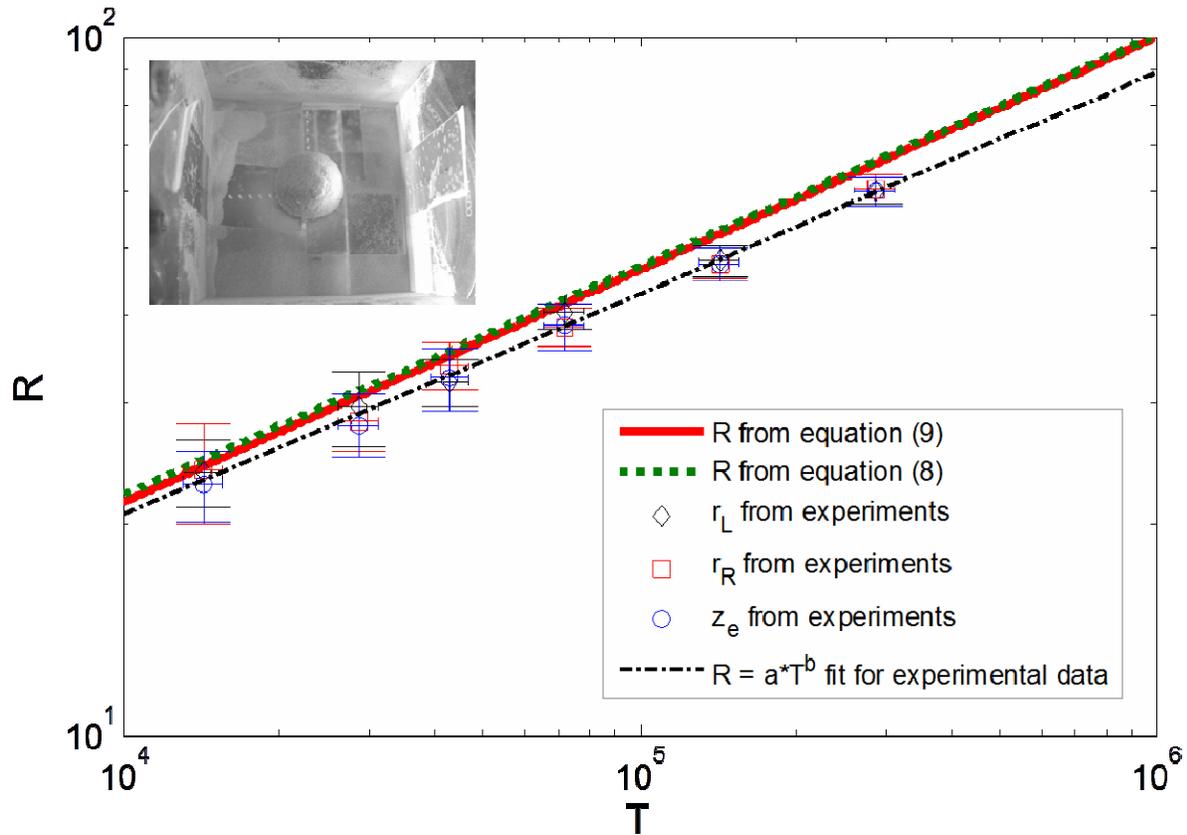

**Figure 4.** Radial growth of an imbibition wetting front in a semi-infinite geometry, measured in three directions as a function of time and compared to the theoretical expression derived in the text. The top left inset shows the container filled with wetted glass microspheres, after suddenly pouring the dry sand out of the container, as described in the experimental section. The liquid is water. Measured values of porosity, permeability and capillary pressure are respectively $\varepsilon = 0.36$, $k = 1.21 \times 10^{-12}$ m$^2$ and $p_c$=3480Pa. Measurements are compared to the full solution of the problem (equation 8) and the approximation for long times (equation 9). The error bars on the measurements reflect the discussion in the experimental section.

The experimental non-dimensional results of position of the front as a function of time are plotted in Figure 4. Three distinct measurements of the front position ($r_L$, $r_R$, $z_e$), as described in the experimental section, are displayed in Figure 4. We chose not to measure data points at times shorter than one minute because the measurements involves turning the box upside down so that the dry sand falls away from the wetted region, which is an operation that takes about 5 seconds. For that reason, data points at shorter times have larger uncertainty than at later times, as can be seen with the vertical error bars. The measurements are non-dimensionalized according to equation (11), assuming that the contact area between the radial source and the porous medium equals the contact area in the experiments, i.e. the cross-section of a tube with diameter $d$. Therefore, $r_s = d/2\sqrt{2}$. The data points for $R_L$, $R_R$ and $R_Z$ overlap at every measured time, suggesting that the front spreads as a hemisphere. The average values fitted on the measurements using the power law $R = a \cdot T^b$ are a=1.03 and b=0.32, respectively. Both the experimental exponent and prefactor are well within 5% of the predicted value of 1 and 1/3 of the power law found analytically in equation (11), respectively, which confirms the cubic root law. The measured slope might be slightly smaller than 1/3 because gravity is not totally negligible for the larger measured radii. Figure 4 also compares the full solution of the imbibition problem, equation (8), with its approximation for long times, equation (9). Results of the comparison show that the relative error between the full solution and its approximation for long times becomes less than 10% for $T > 120$, which in our experiment corresponds to $t > 0.4$ s.

In conclusion, the imbibition process from a point source into a porous material of semi-infinite extent has been studied theoretically and experimentally. These two approaches are in good agreement, and show that when gravity is negligible compared to the capillary pressure, the wetting front conserves a hemispherical shape with radius evolving in time as $r \sim t^{1/3}$. This result complements known one- and two-dimensional imbibition results.

ACKNOWLEDGEMENTS

D.A. acknowledges financial support from the US Department of Justice, Office of Justice Programs, Research and Development on Pattern and Impression Evidence. The China Scholarship Council supported J.X. We thank Patrick Weidman at University of Colorado, and Markus Hilpert at The Johns Hopkins University for sharing thoughts on porous media studies.